\RequirePackage{fix-cm}
\documentclass[twocolumn]{svjour3}          
\smartqed  
\usepackage{graphicx}
\usepackage{amsmath}

\newcommand{\muvec}{\boldsymbol{\mu}}
\newcommand{\Rvec}{\boldsymbol{R}}
\newcommand{\rvec}{\boldsymbol{r}}
\begin{document}

\title{Cold Rydberg atoms for quantum simulation of exotic condensed matter interactions}

\titlerunning{Cold Rydberg atoms and exotic condensed matter interactions}        

\author{J.P. Hague         \and
        S. Downes \and C. MacCormick \and P.E. Kornilovitch 
}


\institute{J.P. Hague, S. Downes and C. MacCormick \at
              Department of Physical Sciences,
              The Open University, Walton Hall, Milton Keynes, MK6 3AY, UK
              \email{Jim.Hague@open.ac.uk}           
              \and
              P.E. Kornilovitch \at Hewlett-Packard Company, 1070 NE Circle Boulevard, Corvallis, Oregon 97330, USA
}

\date{Received: date / Accepted: date}

\maketitle

\begin{abstract}
Quantum simulators could provide an alternative to numerical simulations for
understanding minimal models of condensed matter systems in a controlled
way. Typically, cold atom systems are used to simulate e.g. Hubbard
models. In this paper, we discuss a range of exotic interactions that
can be formed when cold Rydberg atoms are loaded into optical lattices with unconventional geometries; such as long-range
electron-phonon interactions and extended Coulomb like
interactions. We show how these can lead to proposals for quantum simulators for
complex condensed matter systems such as superconductors. Continuous time quantum Monte Carlo is used to compare the proposed schemes with the physics found in traditional condensed matter Hamiltonians for systems such as high temperature superconductors.
\end{abstract}

\section{Introduction}
\label{sec:intro}

The recent implementation of quantum simulators for Hubbard models has
been a triumph of experimental technique \cite{Schneider2008,Jordens2008}. The simulation of Hubbard
models has been a success, however the most basic form of the Hubbard
model misses some crucial physics found in many condensed matter
systems, including: (1) The electron-phonon interaction and (2) Longer
range instantaneous interactions. In this paper we will discuss the
potential for systems of dressed Rydberg atoms to become toolkits for
examining the physics of more complicated, yet more realistic
interactions such as these. Rydberg interactions are particularly
timely, because interactions between small numbers of cold Rydberg atoms have recently been measured
experimentally \cite{schauss2012a}.

Rydberg atoms have been used to propose simulators for various
different types of interactions in condensed matter systems. For
example, chains of Rydberg ions have high energy phonons that can be
used as part of a mapping to spin systems
\cite{porras2004a,mueller2008a}. Other uses of Rydberg states in cold
ion crystals are the investigation of structural distortions
\cite{li2011a}. Simulators for Su-Schrieffer-Heeger polarons relevant
to strongly deformable materials such as polymers may be achievable
using excitons in Rydberg chains \cite{hague2011a}. Cold polar
molecules offer a similar platform to Rydberg states and have been
proposed for simulators of Holstein and Su-Schrieffer-Heeger polarons
\cite{herrera2011a,herrera2013a}. In this paper, the use of Rydberg
atoms to simulate long range electron-phonon couplings for {\it
  arbitrary filling} and extended Hubbard interacions is
discussed. The bilayer scheme discussed here (see also \cite{hague2012b}) has major advantages over
previous proposals for electron-phonon interactions since arbitrary
filling factors could be achieved in principle, whereas previous
proposals concerned only the low density polaron limit. We have not previously discussed extended Hubbard interactions between Rydberg atoms.

\section{Control over Rydberg interactions}
\label{sec:rydberg}

Rydberg atoms have a high principal quantum number. As such, the
electron is far from the nucleus and the atom has a large polarisability. When atoms are in the $ | nS_{1/2}\rangle$ state,
seperated by a displacement $\Rvec$, the dipole-dipole interaction can be calculated by analysing the appropriate interaction:
\begin{equation}\label{eq:DipoleDipolePotential}
V_{\rm dip-dip} =\frac{1}{4\pi \epsilon_{0}}\left(\frac{ \boldsymbol{\mu_{1}} \cdot \muvec_{2}}{R^3}-\frac{3 \left( \muvec_{1} \cdot \Rvec \right)\left( \muvec_{2} \cdot \Rvec \right)}{R^5}\right)
\end{equation}
which couples pairs of $|nS_{1/2};nS_{1/2} \rangle$ states to pairs of other states (e.g. $ | n^{\prime}P_{j^\prime};n^{\prime\prime}P_{j^{\prime\prime}} \rangle$ which differs in energy by $\Delta_{sp}$). Here the dipole moments, $\muvec_{i}=e \rvec_{i}$, with $e$ the electron charge and $\rvec_{i}$ the position operator for atom $i$. The Hamiltonian matrix is then:
\begin{equation}
V_{\rm Int} = 
\left(
\begin{array}{cc}
0 & V(R) \\
V(R) & \Delta_{sp}\\
\end{array}\right)
\end{equation}
where $V(R)=\langle nS_{1/2};nS_{1/2} | V_{\rm dip-dip} |
n^{\prime}P_{j^\prime};n^{\prime\prime}P_{j^{\prime\prime}}
\rangle$. The average over states ensures that $V(R)$ has no
preferential direction. Diagonalizing the matrix leads to the interaction strength,
\begin{equation}
V_{{\rm Int} \pm}= \Delta_{sp}/2 \pm \sqrt{V(R)^2+\Delta_{sp}^2/4}
\end{equation}

If $V(R) \ll \Delta_{sp}$, the standard van der Waals interaction is
recovered. Of more use for condensed matter quantum simulators is the
limit $V(R) \gg \Delta_{sp}$ where the interaction has the form
$V_{\rm Int}=\alpha/R^3$ ($\alpha$ may also be written as $C_3$). This
regime is appropriate for this paper as $\alpha$ is large and
distances in an optical lattice (compared to the size of Rydberg
states) are small. Control over Rydberg states (and the lifetime of
the states) can be improved by illuminating the
atoms with laser light offset by a value $\Delta$ from the transition
between the ground and Rydberg states to form virtual Rydberg atoms.

\begin{figure}
\includegraphics[width=0.47\textwidth]{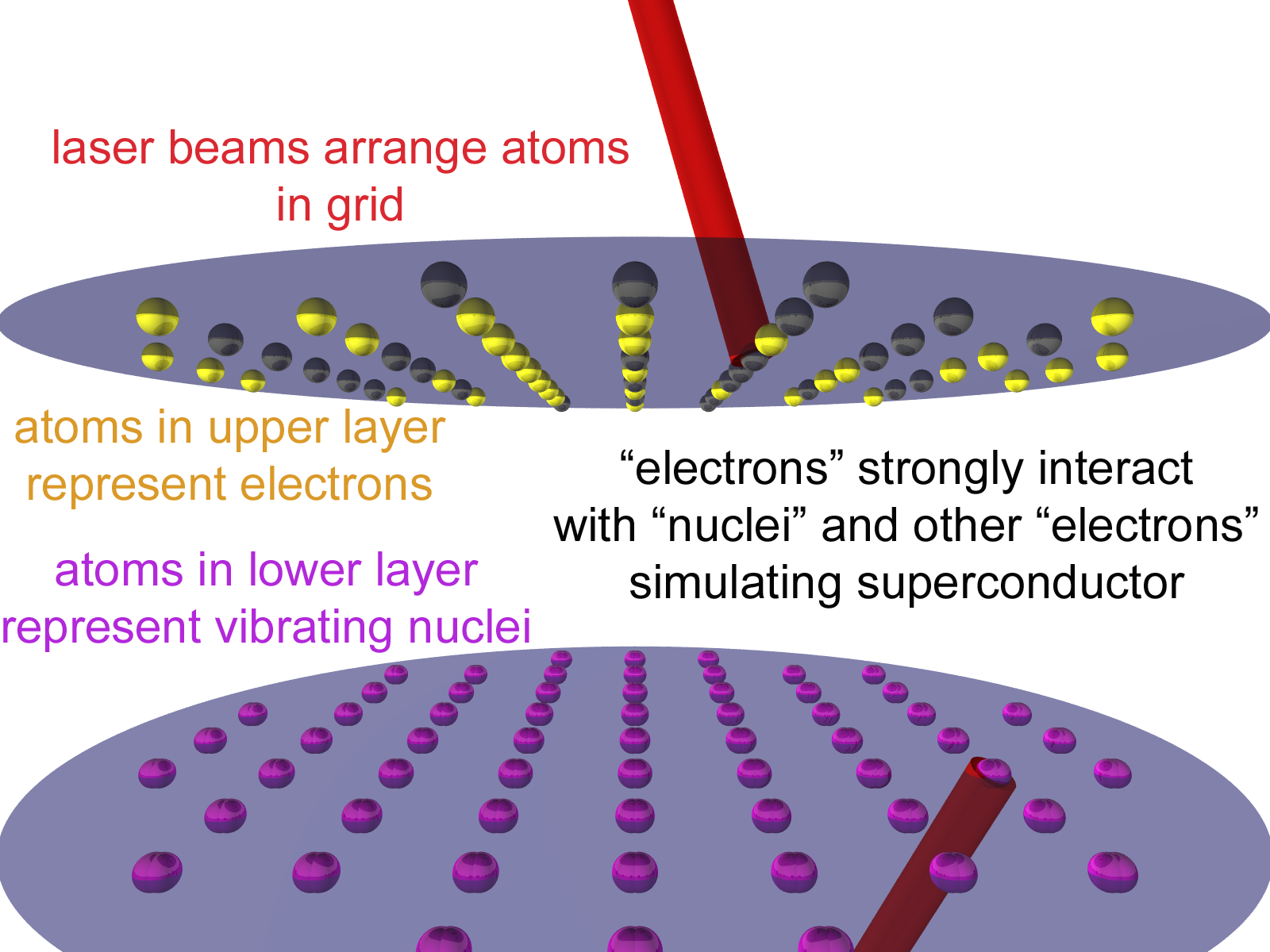}
\caption{Schematic of the bilayer Rydberg system (see main text for more information).}
\label{fig:schematic}
\end{figure}

We propose to load cold Rydberg atoms into a bilayer optical lattice,
similar to that shown in Fig. \ref{fig:schematic}. Each layer
represents different parts of the condensed matter system: Atoms in
the upper layer take the place of itinerant electrons (itinerant
layer), and those in the lower layer vibrating nuclei (phonon
layer). Since each atom in a condensed matter system has a nucleus,
the atoms in the lower (phonon) layer should be in a Mott insulating
state so that there is a single atom (which can vibrate to make phonons) per site - as such the optical
lattice for the phonon layer needs to be deep. The depth of the phonon
layer also means that the phonon layer fills before the itinerant
layer so that it is always completely filled, whereas the itinerant
layer can be partially filled to simulate the effects of doping. In
order to get small phonon frequencies in that layer, two optical
pancakes can be separated by a small distance, $D$, or two spots a distance $D$ apart can
be made using the painted potential technique \cite{Henderson2009}.

The bilayer scheme offers signigicant control over the interactions
and energy scales in the system. The depth of the itinerant layer can
be varied to change energy scales such as the hopping, $t$,
the local Hubbard $U$, the extended Hubbard interaction between atoms
within the itinerant plane, $V_{ij}$. Spots in the phonon layer can be
changed to affect the Rybderg-phonon (electron-phonon) interaction
between planes, $g_{ij}$ and the phonon frequency,
$\omega_{0}$. The strength of $V_{ij}$ and $g_{ij}$ can also be varied by changing the
offset $\Delta$ from the transition when dressing the Rydberg state,
and also by selecting a different prinicipal quantum number of the
Rydberg state. A major advantage of the bilayer scheme over previous
proposals for electron-phonon interactions is that arbitrary filling
factors can be achieved - previously, only proposals for low density
polaron states were available.

Hopping, $t^{(it)}$, Hubbard $U^{(it)}$ and phonon frequencies
$\omega_0^{(it)}$ in the itinerant layer have the standard values
\cite{bloch2008a}:
\begin{equation}
t^{(it)} \approx \frac{4}{\sqrt{\pi}}E_{\rm rec}\left(\frac{V_0}{E_{\rm rec}}\right)^{3/4}\exp\left[-2\left(\frac{V_0}{E_{\rm rec}}\right)^{1/2}\right]
\end{equation}
\begin{equation}
U^{(it)} \approx \sqrt{\frac{8}{\pi}}k\tilde{a}E_{\rm rec}\left(\frac{V_0}{E_{\rm rec}}\right)^{3/4}.
\end{equation}
\begin{equation}
\hbar\omega_0^{(it)}=2E_{\rm rec}\left(\frac{V_0}{E_{\rm rec}}\right)^{1/2}
\end{equation}
where $V_0$ is the depth of the lattice, $E_{\rm rec}$ has the value, $\hbar^2 k^2/2M$, $M$ is the atomic mass of the Rydberg atoms, $k=\pi/a$, $\tilde{a}$ is the s-wave scattering length and $a$ is the lattice constant. Both $t^{(it)}$ and $U^{(it)}$
may be adjusted to represent hopping and Hubbard $U$ in the resulting
electron-phonon problem. Note that the phonon frequency $\omega_0^{(it)}$ is not
the phonon freqency in the simulated problem, rather in this case it
is only important that $\omega_0^{(it)}$ is very large as it essentially
represents the separation between bands in the problem (energy levels
of the simulated electron).

The phonon frequency for the electron-phonon Hamiltonian is provided by the phonon layer. The potential provided by the spot potentials is:
\begin{equation}
-\frac{V^{(ph)}_0}{2}\left[\exp\left(-\frac{(x-D)^2}{2w^{(ph)^2}}\right)+\exp\left(-\frac{(x+D)^2}{2w^{(ph)^2}}\right)\right]
\end{equation}
(where the depth of the potential in the phonon layer is $V_0^{(ph)}$ and $w^{(ph)}$ is the waist size of the beam forming the painted potential, or in the case of 2 slightly separated optical pancakes is the width of each pancake). The Taylor expansion leads to the phonon frequency,
\begin{equation}
(\omega_{0}^{(ph)})^2=V_0^{(ph)} \exp\left(-\frac{D^2}{2 w^{(ph)^{2}}}\right) \frac{(w^{(ph)^{2}} - D^2)}{M w^{(ph)^{4}}}.
\end{equation}

Taylor expansion of the dipole-dipole interactions between Rydberg
atoms in itinerant and phonon layers and assuming only c-axis phonons
(which can be achieved by making the phonon layer from 2 optical
pancakes separated by a small distance, $D$, as with the phonon
frequency) leads to an electron-phonon (Rydberg-phonon) interaction
$g_{ij}$:
\begin{equation}
g_{ij} = \frac{\Omega^2}{4\Delta^2}\frac{3\mu^2 b}{(b^2+r_{ij}^2)^{5/2}}\sqrt{\frac{\hbar}{2M\omega_0}}
\label{eqnrydphonon}
\end{equation}
where $\Omega$ is the Rabi frequency and $b$ is the interplane distance.

Taking these values together, a dimensionless electron-phonon coupling
can be defined,
\begin{equation}
\lambda = \frac{1}{2\omega_0^{(it)2} M W}\left[\frac{3\Omega^2\mu^2}{4\Delta^2}\right]^2\sum_i\frac{b^2}{(b^2+r_{0i}^2)^{5}}
\end{equation}
where $W=4t^{(it)}$.

Finally, it should be noted that there are dipole-dipole interactions between Rydberg atoms in the itinerant plane that contribute an extended Hubbard V to the Hamiltonian,
\begin{equation}
V_{ij} = \frac{\left(2M\omega_0^{(it)2} \lambda W \right)^{1/2}}{3r_{ij}^{3}}\left[\sum_{k} \frac{b^2}{(b^2+r_{0k}^2)^{5}}\right]^{-1/2}.
\end{equation}
In the limit of low phonon frequency, this does not contribute to the
physics as demonstrated elsewhere \cite{hague2012b}. Alternatively, it
is possible to manipulate the Rydberg system so that this extended
Hubbard physics dominates, and in this case we use the shorthand $V_{ij}=\alpha/r_{ij}^3$.

Thus, the full Hamiltonian that is simulated by the proposed bilayer Rydberg quantum simulator is:
\begin{eqnarray}
H & = & -t^{(it)}\sum_{\langle ij \rangle}c^{\dagger}_{i}c_{j} + U^{(it)}\sum_{i}n_{i\downarrow}n_{i\uparrow} + \sum_{i\neq j}V_{ij}n_{i}n_{j}\nonumber\\& & + \sum_{ij} g_{ij}n_{i} (d_{j}+d^{\dagger}_{j}) + \hbar\omega^{(ph)}_0 \sum_{i}d^{\dagger}_i d_i
\end{eqnarray}


\section{Tests of the proposed simulator}
\label{sec:results}

\begin{figure}
\includegraphics[width=0.47\textwidth]{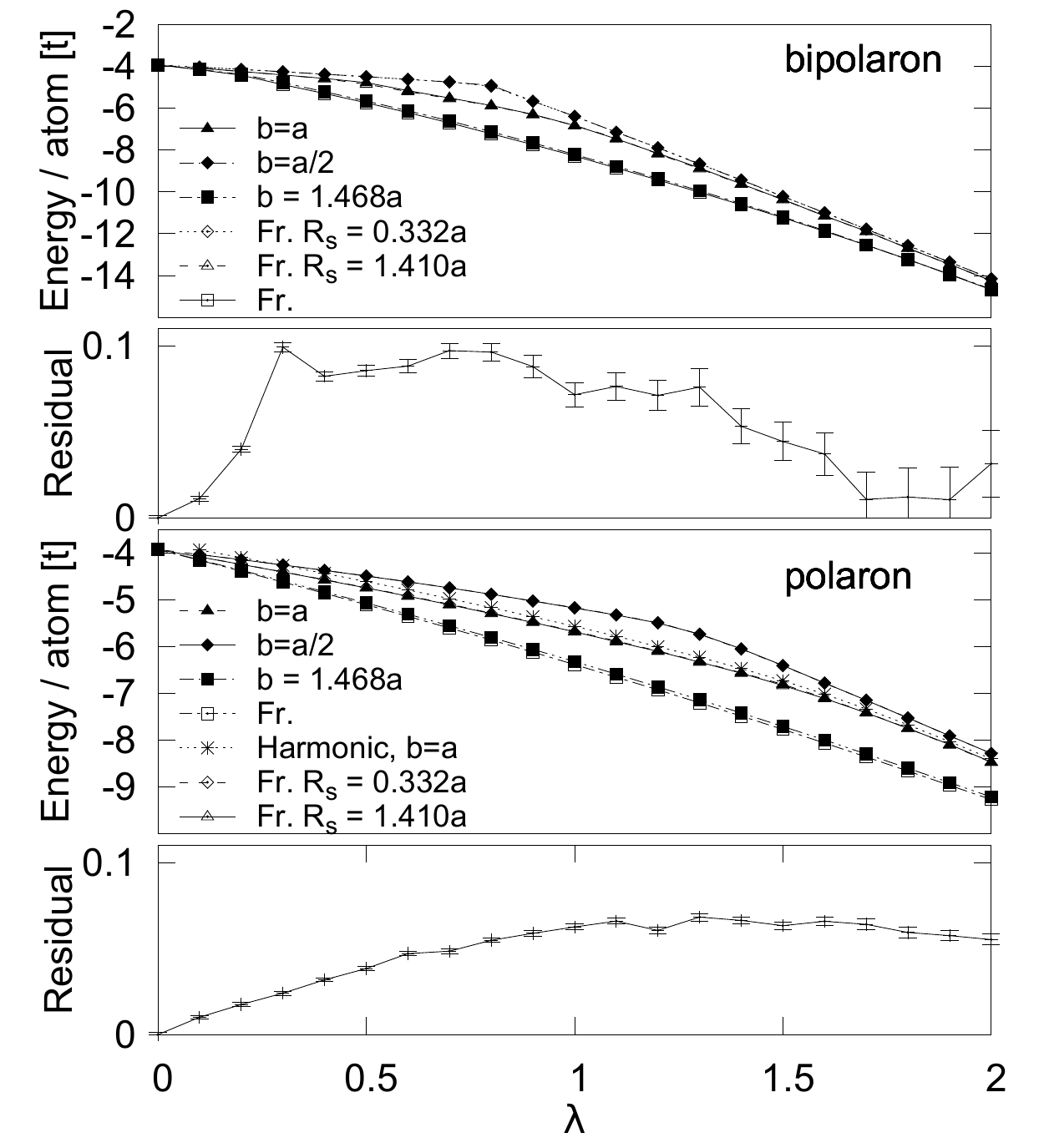}
\caption{Polaron and bipolaron energies for different interaction ranges within the bilayer Rydberg scheme proposed here, and for a
  screened Fr\"ohlich interaction. The near-neighbor interaction
  strengths have been matched. $U^{(it)}=4t^{(it)}$,
  $\hbar\omega^{(ph)}_0=t^{(it)}$, $\lambda =
  \Phi(0)/8t^{(it)}M\omega_0^{(it)2}$. Unless shown, error
  bars are smaller than the points. The lower plots show the
  residual between the Rydberg and Fr\"ohlich schemes, which is less than $0.1t$ per particle. Both the interaction type in the proposed quantum simulator and the screened Fr\"ohlich interaction have quantitatively similar physics. The effects of a harmonic potential are also shown, and do not lead to significant deviations from the free lattice case.}
\label{fig:polaron}
\end{figure}

The Rydberg-phonon interaction in Eqn. \ref{eqnrydphonon} differs
slightly from the standard Fr\"ohlich form of the electron-phonon
interaction,
$g^{(Fr)}_{ij}\propto\exp(-R_{sc}r_{ij})(a^2+r_{ij}^2)^{-3/2}$. In the limit
that $b\rightarrow 0$, the standard Holstein interaction
$g^{(Hol)}_{ij}\propto\delta_{ij}$ is recovered. It is necessary to check that
the difference in the tails of the interaction does not lead to any
major qualitative changes in the physics of the Hamiltonian. Here,
continuous time quantum Monte Carlo (CTQMC) \cite{hague2009a} is used to examine
the polaron and bipolaron physics, as shown in
Fig. \ref{fig:polaron}. The figure shows results from QMC for both Fr\"ohlich
interactions, and the Rydberg-phonon interaction considered
here. Effective nearest-neighbour interactions have been matched by
modifying the distance $b$ between the two planes, so only the tails of the interactions differ as would be the case when making an experimental implementation of the scheme. The agreement
between the two schemes is excellent.

\begin{figure}
  \includegraphics[width=0.47\textwidth]{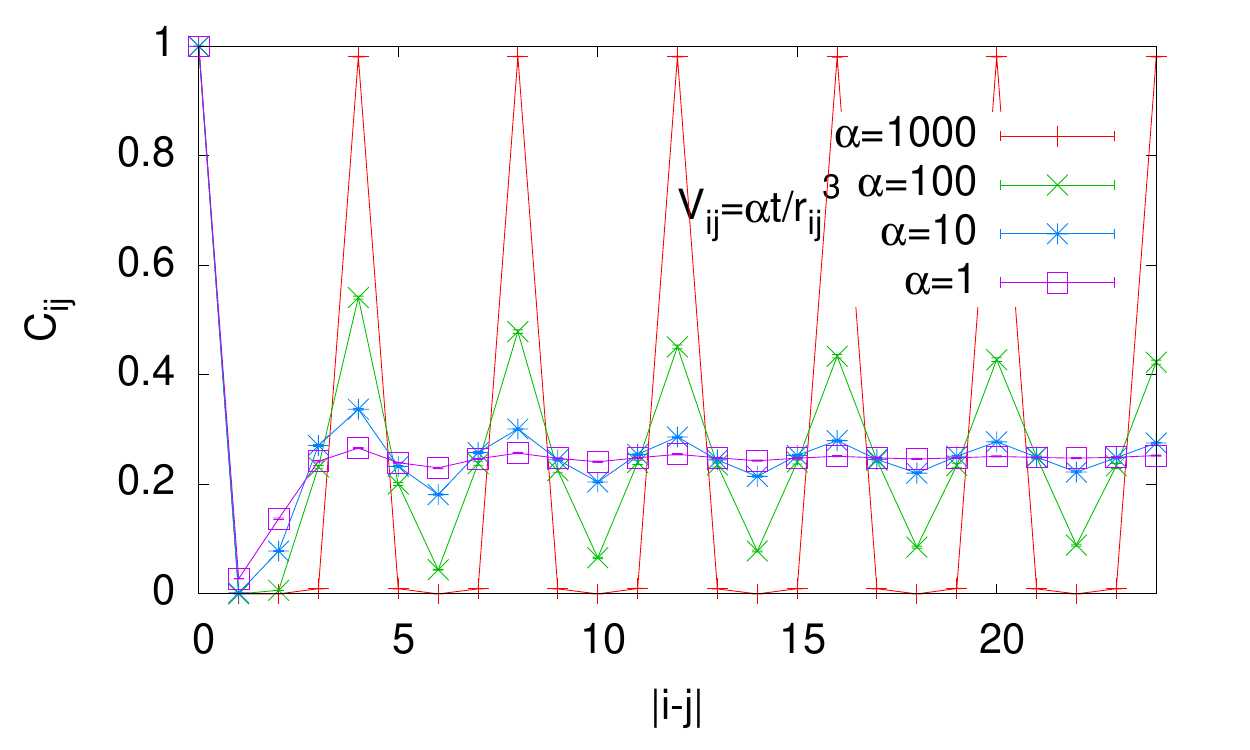}
\caption{Density-density correlation function vs relative particle positions, for a range of interaction strengths, $\alpha$. For large $\alpha$, the system forms a state with local order. As $\alpha$ is decreased, the long range behaviour becomes unordered (see the long tail of $C_{ij}=0.25$ when $\alpha=1$). Error bars are typically too small to be distinguishable.}
\label{fig:correlator}
\end{figure}

As yet unexplored is the potential to use Rydberg atoms for extended
Hubbard interactions. A 1D simplification of the complex bilayer
geometry allows the simulation of many particle ensembles using CTQMC
without sign problems. Fig. \ref{fig:correlator} shows the correlation
function of particle positions $\langle n_{i}n_{j}\rangle$ (including
auto-correlation). 12 particles were simulated on a ring of 48 sites
(density 1/4). For large $\alpha$, short range correlations can be
observed as a modulation of the correlation function. As $\alpha$
decreases, the range of the correlations decreases until the tails of
the correlation function show the uncorrelated value of
$0.25$. Correlations and extended Hubbard interactions from Ryberg
atoms will be described in more detail in an extended paper.

\section{Summary and outlook}
\label{sec:summary}

We have proposed that systems of Rydberg atoms could be used to simulate
exotic long-range interactions in condensed matter systems. States of
dressed Rydberg atoms can be chosen such that the dipole-dipole
interactions between them have the form $1/R_{ij}^3$. Appropriate
layering of Rydberg atoms into bilayers where an itinerant layer
represents the physics of itinerant electrons and the other layer
represents vibrating nuclei (phonons) leads to Hamiltonians including
interaction terms analogous to the Hubbard U, extended Hubbard
$V_{ij}$, long range electron-phonon interaction, low frequency
phonons and inter-site hopping, $t$. We have previously calculated that this is possible with current cold atom technology \cite{hague2012b}. Tests using continuous time QMC simulations
have shown that the physics of the Rydberg-phonon interaction is
similar to the lattice Fr\"ohlich electron-phonon interaction with very good agreement in the physics of both interactions. Further
simulations on a simplified ring geometry have shown (local) ordered
states from an extended Hubbard interaction. These states will be
discussed in more detail in an extended paper.


\bibliographystyle{spphys}       
\bibliography{rydtool.bib}   

%
%

\end{document}